8 January 2011

# Quantum control in spintronics

A. Ardavan

Department of Physics, University of Oxford, The Clarendon Laboratory, Parks Road, Oxford OX1 3PU

G.A.D. Briggs

Department of Materials, University of Oxford, Parks Road, Oxford OX1 3PH

*Abstract*

Superposition and entanglement are uniquely quantum phenomena. Superposition incorporates a phase which contains information surpassing any classical mixture. Entanglement offers correlations between measurements in quantum systems that are stronger than any which would be possible classically. These give quantum computing its spectacular potential, but the implications extend far beyond quantum information processing. Early applications may be found in entanglement enhanced sensing and metrology. Quantum spins in condensed matter offer promising candidates for investigating and exploiting superposition and entanglement, and enormous progress is being made in quantum control of such systems. In GaAs, individual electron spins can be manipulated and measured, and singlet-triplet states can be controlled in double-dot structures. In silicon, individual electron spins can be detected by ionisation of phosphorous donors, and information can be transferred from electron spins to nuclear spins to provide long memory times. Electron and nuclear spins can be manipulated in nitrogen atoms incarcerated in fullerene molecules, which in turn can be assembled in ordered arrays. Spin states of charged nitrogen vacancy centres in diamond can be manipulated and read optically. Collective spin states in a range of materials systems offer scope for holographic storage of information. Conditions are now excellent for implementing superposition and entanglement in spintronic devices, thereby opening up a new era of quantum technologies.

## 1. The full quantum potential of spin states

Electron spin is by its nature quantum in origin. In that sense all manifestations of spin, such as magnetism, are quantum, and technologies which use spin, such as spintronic valves for read-heads, are also quantum. But there is a further resource in superposition, the uniquely quantum phenomenon whereby something can exist simultaneously in more than one state, for example in the case of an electron spin, both up and down. Superpositions involving two or more spins can exhibit entanglement, with correlations between measurements that exceed anything which could be accounted for by classical physics. It is superposition and entanglement that distinguish quantum technologies and give them their spectacular potential. Superposition and entanglement of electron spin states have been widely studied in the quest to build a solid state computer. They have yet to be implemented and exploited in practical spintronics. Here, we describe some of the contributions which we and our immediate colleagues have made to the field, together with other selected ground-breaking experimental advances towards quantum spintronics technologies. This article is not a comprehensive review, but is intended to impart some of the excitement of the field to those working in "classical" spintronics.

## 2. Quantum information in electron and nuclear spins

Classical binary information may be stored in any system with two distinguishable states; these states may be (and usually are) macrostates. An element of quantum information, in contrast, is represented by the superposition of two quantum states, each of which is necessarily a microstate of a system. Any physical system that is to be exploited as a quantum information processor must therefore exhibit a set of quantum two-level systems (qubits), whose states may be manipulated by the "operator" of the quantum computer. A further requirement is that conditional logic should be possible, that is, the state of one qubit may be manipulated in a way that depends on the state of another qubit. Quantum spins were identified early as candidates for embodying qubits; the two-level structure of a quantum spin-½ naturally forms a qubit, and couplings between spins offer mechanisms for conditional logic.

Early in the development of experimental quantum information research, it was recognised that many of these requirements were already satisfied in nuclear magnetic resonance (NMR) experiments.[1] The magnetic nuclei in a molecule have distinct NMR frequencies by virtue of their different moments and chemical environments, and may therefore be individually manipulated using radio frequency pulses of distinct frequencies. The effective Ising interaction between nuclei, mediated by the electronic molecular orbitals, may be exploited for conditional manipulations. A culmination of these experiments was the factorisation of 15 using Shor's algorithm.[2] Shor's algorithm has been hugely influential in motivating quantum computing research, because through its ability to factor the products of prime numbers it offers the prospect of codebreaking.[3]

There are serious limitations to NMR quantum computing that prevent scaling it up to a number of qubits sufficient to solve problems that could not more easily be solved on a conventional computer.[4] First, what is detected in an NMR experiment is the macroscopic magnetisation of a large ensemble of molecules (it is, as yet, not possible to detect the states of individual nuclei), so that the calculation is performed simultaneously on a very large number of copies of the quantum computer. Second, in any practical NMR experiment the degree of polarization of the spins is very weak. In combination, these factors introduce fundamental constraints on achieving entanglement, and practical constraints on the ability to measure the signal (which decreases exponentially as the number of qubits grows).

Electron spins have a magnetic moment which is three orders of magnitude larger than nuclear spins. This offers correspondingly higher polarization even at room temperature, and at experimentally accessible combinations of field and temperature the polarization can approach 100%. It also offers routes to the detection of the states of individual qubits.

## 3. Single electron spin manipulation in double dots in compound semiconductors

If an electron spin is to be useful as a qubit, it is helpful for it to be localised (though, as we shall see later, collective non-local states of many particles can also embody qubits). In an influential proposal for quantum information processing in the solid state, Loss and DiVincenzo suggested a scheme building upon extensive research into electrostatically defined quantum dots in GaAs-AlGaAs heterostructures.[5] In these systems, electrons which are already confined to a 2D layer at the heterostructure interface are further confined by potentials applied using electrostatic gates on the surface.[6] Electron densities in the 2D layer are typically $10^{15}$ m$^{-2}$, and the gate structures are defined using electron beam lithography with typical dimensions of a few tens of nanometres, resulting in islands containing very small numbers of, or even single, electrons.

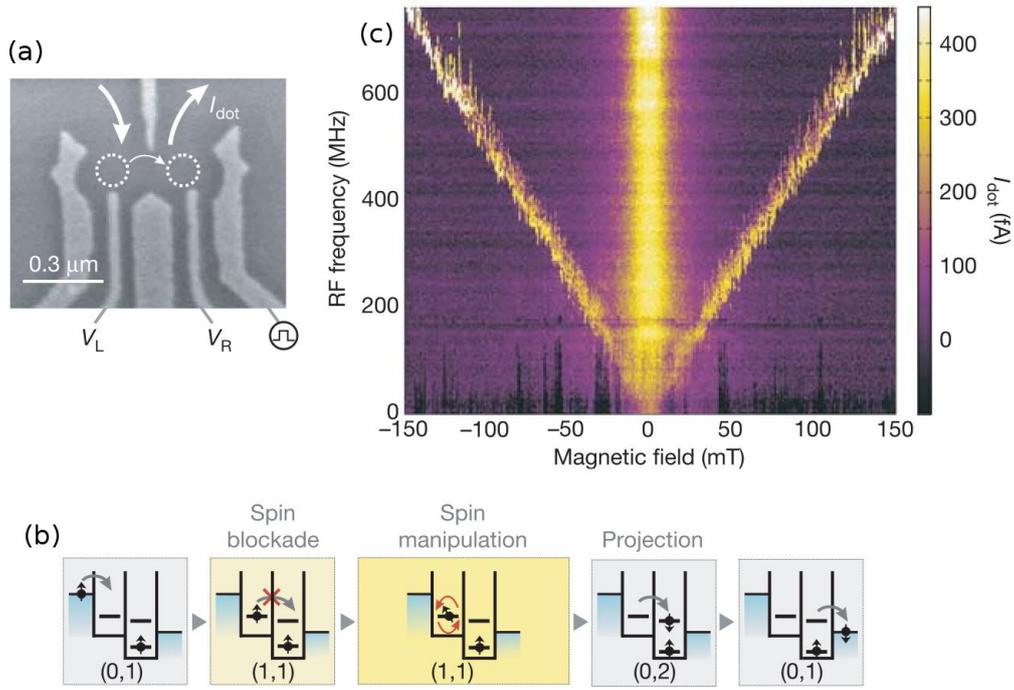

Figure 1: (a) Micrograph of a gate confined double dot device; (b) Schematic of the process used to prepare, manipulate and detect the spin state of individual electrons in the double dot; (c) Current through the double-dot system as a function of magnetic field and frequency of applied microwave field. Electron spin resonance lifts the spin blockade, increasing the current through the dots. Adapted from Ref. 9.

Experiments are typically performed on double-dot devices, an example of which is shown in Figure 1(a). The six independently controllable gates allow independent tuning of the key parameters of the double dot system: the electrochemical potential of each dot; the coupling (or tunnelling rate for electrons) between each dot and the surrounding 2D electron gas (2DEG); and the coupling between the two dots. By studying the behaviour of the conductance through the double-dot structure as a function of the gate voltages (revealing the famous Coulomb-diamond patterns[6]) each of these parameters can be measured, and the regime of interest for quantum information processing can be identified.[7] Usually, this regime corresponds to the range of gate voltages for which there are one or two conduction-band electrons in the entire structure. Exploiting the Coulomb energy associated with the occupation of one dot by two electrons, the exchange energy associated with the coupling between the two dots, and the independently tuneable electrochemical potential of each dot, and working at temperatures at which $k_B T$ is small compared to these energy scales, it is possible to manipulate and study the occupations and spin states of electrons in the dots in extraordinary ways.

One of the key experiments demonstrating such control was the detection of spin resonance in an individual electron in a double-dot structure. The scheme is illustrated in Figure 1(b). The gates are prepared so that the right-hand dot contains a single electron, and an external magnetic field is applied ensuring that its spin state is up. The electrochemical potential of the left dot is below that of the 2DEG to the right, so an electron tunnels onto the left dot. If its spin is down (i.e. it forms a singlet state with the electron in the right dot), it can tunnel onto the right dot and out into the right-hand 2DEG. However, if its spin is up it cannot do so, because the energy of a triplet on the right dot is much higher. This is known as the "spin-blockade regime", in which the charge state of the dots and the current through them are dependent on the spin states of the electrons. Since single-electron charges are much easier to detect than single-electron magnetic moments, this forms the basis for a technique for measuring the relative orientation of the spins of the two electrons in the dot. This is an example of a class of techniques known as "spin to charge conversion".[8]

In the experiment depicted in Figure 1, a further ingredient was the addition of a strip line capable of exposing the double dot structure to microwave radiation. The system is prepared as described above with one electron in each dot, in a triplet state. If the microwave field is resonant with the Zeeman transition of one of the electrons (i.e. it stimulates spin flips), the spin blockade is lifted and a current can be detected. As shown in Figure 1(c), there is a peak in current through the structure as a function of magnetic field and microwave frequency, demonstrating electron spin resonance of individual electrons. By pulsing the gate voltages and the microwave field it is possible to rotate the state of one of the electrons in a controllable way, so that the system exhibits Rabi oscillations.[9]

Using microwave magnetic fields to perform spin manipulations has significant disadvantages, particularly when applied to nanoscale devices. It is difficult to localise the oscillating magnetic field, typically generated by a nearby oscillating current, to the particular quantum dot of interest, and its effect on other components of the device nearby must be considered. In practice, this implies that all nearby elements of the device must be detuned during the microwave pulse so that only the component on resonance is affected. However, spatially localising an oscillatory electric field is more straightforward. In a development of the experiment shown in Figure 1, modulating the voltage on the gate confining the right-hand dot at microwave frequencies induces an effective oscillatory magnetic field via the spin-orbit interaction.[10] Thus the oscillatory electric field can excite Zeeman transitions, offering a convenient local means of manipulating spins in quantum dots.

## 4. Singlet-triplet qubits in double dots in compound semiconductors

Experiments with an alternative double-dot geometry are illustrated in Figure 2. As before, the dots are defined by surface gates above a GaAs-AlGaAs heterostructure, but detection of the state of the system is achieved using a quantum point contact (QPC) defined close to the right-hand dot (Fig. 2(c)). The conductance of the QPC depends on the charge configuration in the double dot, so the experiment probes the charge state of the system directly, rather than via the current through the structure.

Double-dot structures of this kind afford remarkable control over the spin states of two electrons, and have been used to explore the possibility of using singlet and triplet states for embodying quantum information. The coupling of spins of individual electrons to magnetic fields offers a means of manipulating the spin state, but also causes loss of the quantum information stored if the environment exhibits uncontrolled fluctuating magnetic fields. However, if two similar electrons are available for storing each qubit, it is possible to identify states of the system that are immune to magnetic field fluctuations; the two quantum levels representing the "logical qubit" are the singlet state (S) and the triplet state with zero projection in the direction of the external applied magnetic field ($T_0$). The energies of these two states respond equally to fluctuations in the magnetic field, and this is known as a "decoherence free subspace".[11]

One of the key experiments implementing these ideas is the demonstration of coherent control of a singlet-triplet logical qubit. Using a device similar to that shown in Figure 2(c), the system is prepared in a state with one spin down and one spin up. This is an equal superposition of the singlet and triplet states, and so represents a superposition state of the qubit. Controlling the exchange interaction between the two spins, i.e. the singlet-triplet energy splitting, provides the means for controlling the qubit state. Experimentally, this is achieved by manipulating the voltages on the gates confining the electrons.[12] Using elegant techniques derived from pulsed magnetic resonance to overcome the effects of random but slowly varying effective magnetic fields arising from the magnetic nuclei in the vicinity of the electron spins, the authors demonstrated storage and manipulation of quantum information in a two-electron logical qubit over microsecond timescales.[12]

8 January 2011

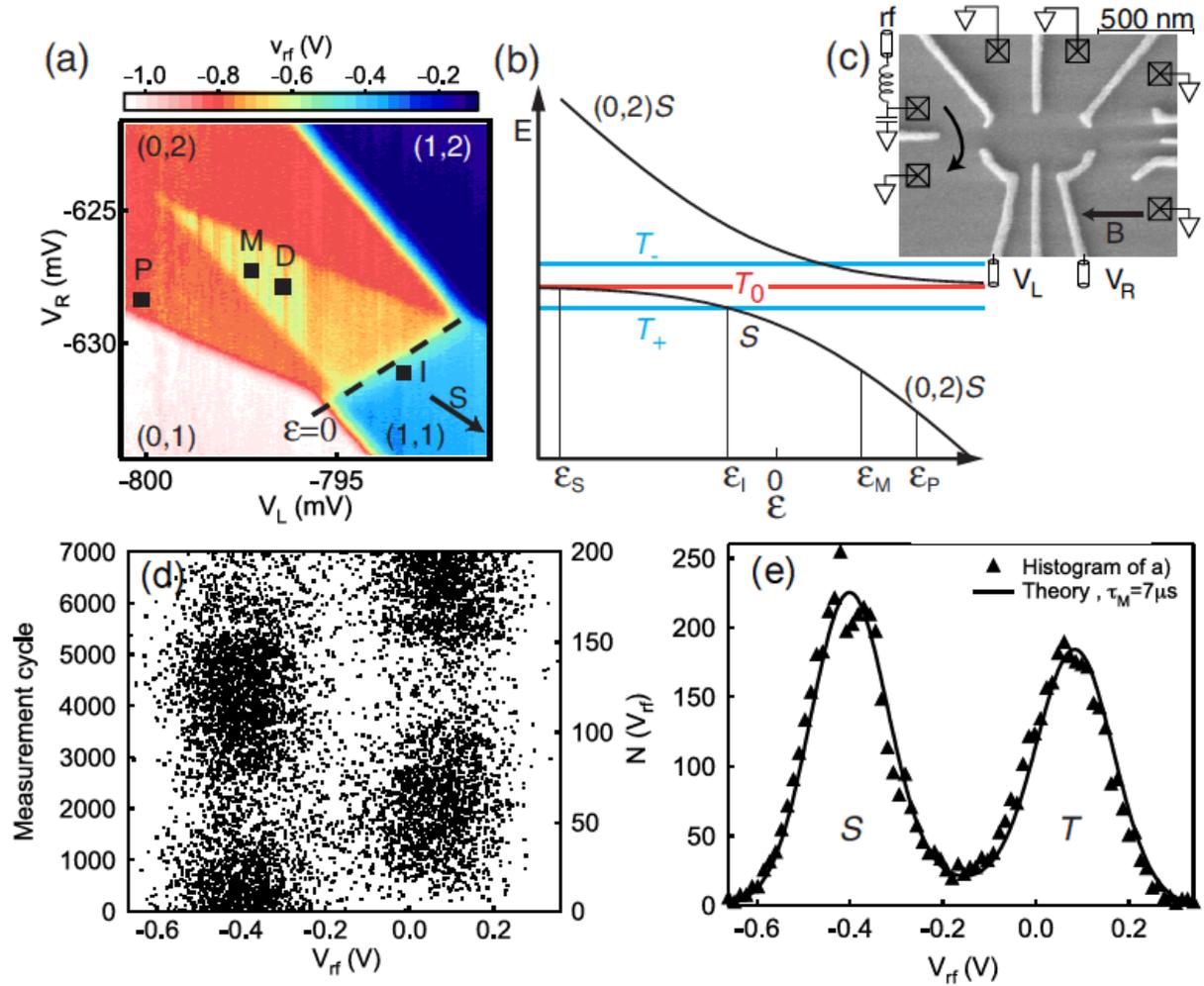

Figure 2: (a) The "Coulomb diamond" stability diagram. The charge state (left and right dot occupancies indicated by numbers in parentheses) is detected using the high frequency QPC, as a function of the voltages applied to the confining gates, $V_L$ and $V_R$, as shown in (c). (b) The energies of the singlet and triplet configurations as a function of the "detuning" $\varepsilon$, the difference between $V_L$ and $V_R$. (d) The results of single-shot measurements of the spin state of the two-electron qubit via the high frequency QPC. (e) The distribution of voltages measured across the high-frequency QPC, showing peaks at -0.4 V representing the singlet state, and at +0.1 V representing the triplet. Adapted from Ref. 13 courtesy C. M. Marcus.

In these experiments, the measurement of the charge state via the conductance of the QPC is slow, which means that the experiment must be repeated many times to accumulate a detectable signal. A spectacular recent development is to operate the QPC in a high frequency mode, enabling measurements on the microsecond timescale.[13] This makes it possible to perform single-shot measurements distinguishing the singlet and triplet states of the logical qubit. Figures 2(d) and 2(e) show the output of many such experiments; the measurements of the output of the high frequency QPC are clustered around two values (-0.4 V representing the singlet state, and +0.1 V representing the triplet). As a function of time the system evolves between the singlet and triplet states. Usually, Rabi oscillations of this kind are presented as the smooth sinusoidal variation of some expectation value derived from a multi-shot or ensemble averaged experimental measurement. In contrast, the true single-shot nature of the detection of this Rabi oscillation is manifest in the form of Figure 2(d); each projective measurement yields either singlet or triplet. Using increasingly sophisticated pulse sequences, it has proved possible to extend the coherence times in these materials to 80 $\mu s$[14] or even 200 $\mu s$.[15]

## 5. Dopant spins in silicon

A second influential proposal for solid state quantum computing was made by Kane.[16] This scheme, in its original form, uses the nuclear spins of phosphorous donors in silicon (I = 1/2) as the qubits, with the spins of the bound electrons used for control, interactions and read out. The interaction between electron spins and individual nuclear spins is controlled by one kind of electrostatic gate (known as an A gate), and the displacement of electrons to mediate an interaction between two nuclear spins is controlled by another class of gates (the J gates). Although Kane's proposal employs materials and dopants that are already widely deployed in current information technologies, the construction of a Kane computer presents formidable challenges in the placing of individual phosphorous donor atoms with nanometre precision in crystalline silicon, and in read out of single electron spins. Experiments have been performed on resonant tunnelling devices fabricated from a very small number of controllably implanted phosphorus donors in silicon, demonstrating the control over the charging of individual donor sites and the Zeeman splitting of the spin states of donor-bound electrons.[17]

A significant development in this area is the single-shot readout of the spin state of electrons apparently bound to single phosphorus donor atoms.[18] About three phosphorus atoms were implanted a few tens of nanometres from a gate-defined single-electron transistor (SET) quantum dot, in a device like the one shown in Fig. 3(a). The electrochemical potential of the SET island is tuneable into a regime in which the current through it is sensitive to whether or not the nearby donors are ionized. The proximity of the donors to the SET island ensures that the electrons can tunnel between the island and the donors on a timescale ranging from microseconds to seconds. A gate situated above the implanted donors tunes the electrochemical potential of the donors with respect to that of the SET island, allowing controlled loading and unloading of electrons to and from the donors. A magnetic field of a few Tesla splits the spin states of the occupied donors by more than the thermal energy at the temperature at which the experiment was performed, which was an electron temperature of order 200 mK.

Experimental cycles are illustrated in Figs 3(b) and (c). First there is a load stage, when the gate voltages are held such that the donor electrochemical potential is well below that of the SET island, and an electron tunnels from the SET island onto the donor with a high probability. Second there is a read stage, when the donor electrochemical potential is raised with respect to that of the SET, with the intention of allowing the electron to escape from the donor depending on its spin state; if the electron does tunnel, the SET turns on. Figure 3(b) illustrates the case where the spin is down, and no read pulse is obtained. Figure 3(c) contains a single current pulse at the beginning of the read phase, which indicates a spin up state. Finally there is reset stage where the donor is emptied ready for the next cycle.

Evidence that the electron is indeed occupying a donor level during the load phase is offered by a measurement of the spin relaxation time $T_1$. An electron that is loaded onto a donor arrives with a random spin, so if its spin is measured immediately using the procedure described above, the spin-up probability is up to 50%. However, if it is held on the donor for an interval comparable with the relaxation time, the spin-down probability increases, as illustrated for different magnetic fields in Fig. 3(d). This provides a mechanism for probing the spin relaxation time, and its magnitude and magnetic field dependence are found to be remarkably similar to traditional electron spin resonance studies for bulk phosphorus-doped silicon.[19] The read out fidelities for spin up and spin down as a function of threshold current are shown in Fig. 3(d), together with the readout visibility. The next step will be to demonstrate coherent control of the trapped electron, by manipulating it either with electric or magnetic oscillatory fields, and a measurement of the phase relaxation time $T_2$. This would also enable an experiment to confirm that the electron is bound to a phosphorus atom, by detecting the hyperfine coupling with the $I = ½$ phosphorus nucleus.

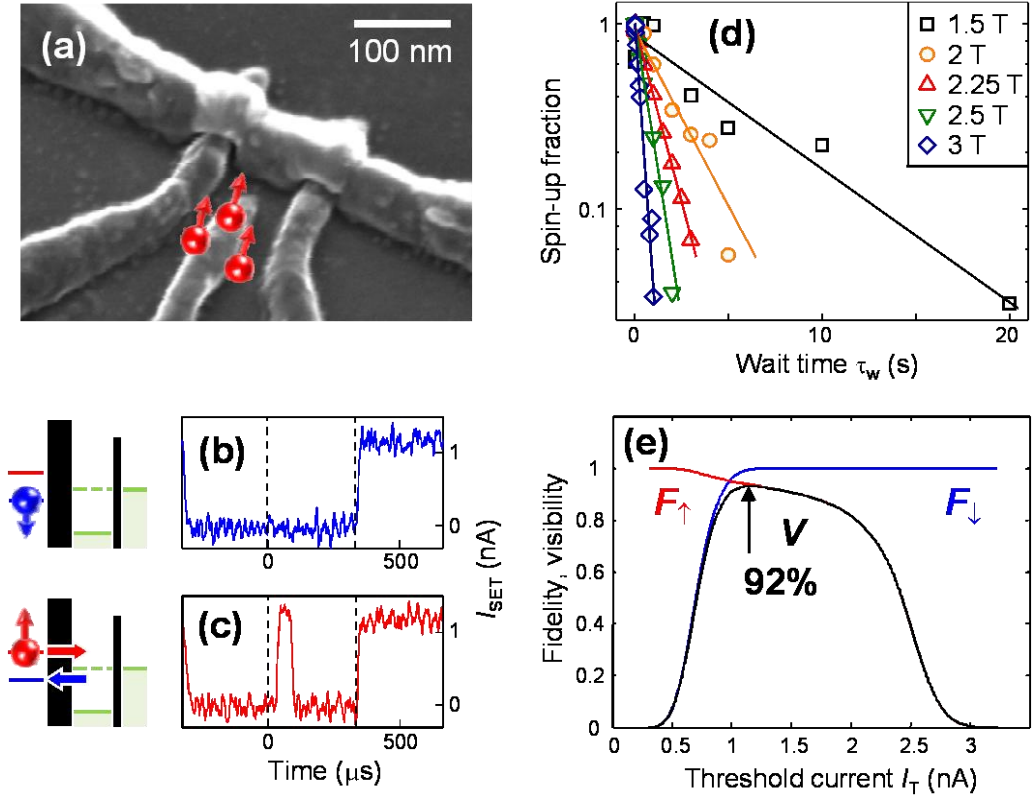

Figure 3: Single-shot readout of a single spin in silicon. (a) Scanning electron micrograph of a device similar to the one measured. The implanted P donors are drawn symbolically in red. (b) and (c) SET single-shot traces for spin down and spin up, with an applied magnetic field B = 5 T. (d) Exponential decays of the normalized spin-up fraction at different magnetic fields. (e) Spin up (blue) and spin down (red) readout fidelities, and readout visibility (black) as a function of the discrimination threshold, $I_T$. The maximum visibility is 92% at $I_T \approx 1.1$ nA. From Ref. 18.

## 6. Vital statistics of molecular electron spins

Quantum superpositions in electron spin qubits can be controlled with exquisite precision using magnetic resonance techniques. By applying microwave pulses of chosen duration, amplitude, and phase, the quantum states of spins can be manipulated at will. Imperfections in the control pulses can be compensated by using sequences originally developed for nuclear magnetic resonance, which can reduce the effect of small systematic errors in the pulses to their sixth power in single qubit gate operations.[20,21] The dominant isotopes of carbon and of silicon have no nuclear spin, and therefore these elements can support electron spin qubits with remarkable coherence times.

The fullerene family offers spins suitable for quantum computing by incarcerating one or more spin-bearing atoms to make endohedral fullerene molecules.[22] These molecules lend themselves to chemical assembly, and they can even be arranged in linear arrays in single-walled carbon nanotubes,[23] with gates for addressing qubits and controlling interactions.[24] With the development of aberration corrected transmission electron microscopy at low voltage (80 kV), to minimise knock-on damage, it has become possible to image the actual piece of active material in a device, with resolution considerably finer than the spacing between the carbon atoms in the nanotube,[25] and even observe nanoscale motion.[26] Figure 4 illustrates how these materials can be imaged, showing pictures of a nanotube and of a so-called peapod of $La@C_{82}$ endohedral fullerene molecules in a single-walled carbon nanotube.[27]

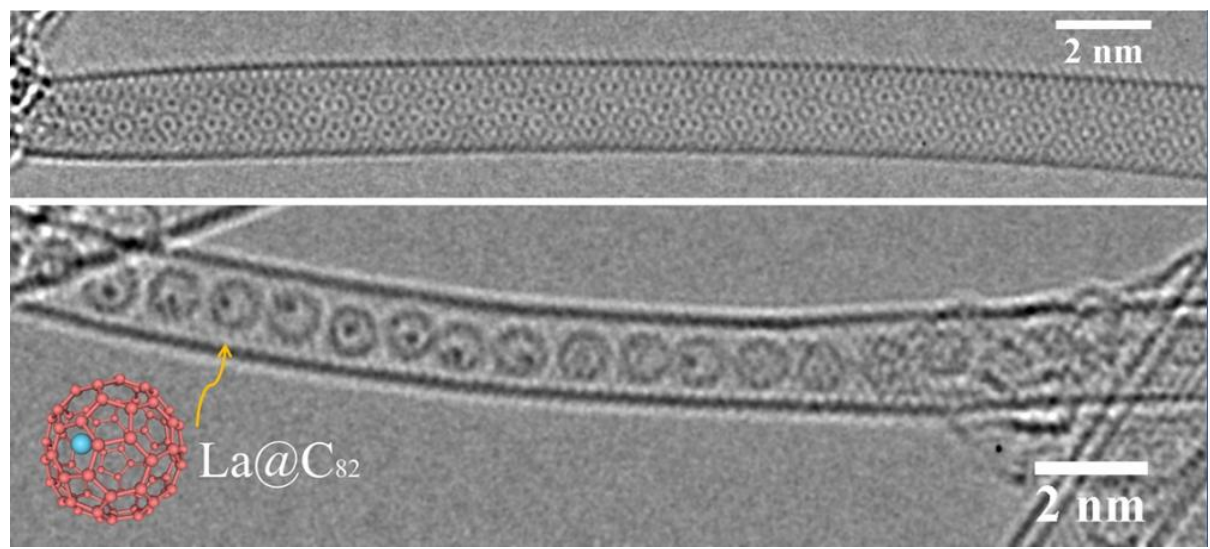

Figure 4. *Upper*: a single walled nanotube; *lower*: a La@$C_{82}$ peapod. The structures are observed in transmission electron microscopy (aberration-corrected JEOL 2200MCO operating at 80 kV, images courtesy of Dr Jamie Warner).

The endohedral fullerene with the longest electron spin coherence time consists of a single nitrogen atom in a cage of sixty carbon atoms, N@$C_{60}$.[28] This molecule is something of a miracle, since a nitrogen atom is chemically reactive, and might be expected to bond strongly to one or more of the surrounding carbons atoms. Astonishingly, the nitrogen atom sits in the middle of the fullerene cage, with its electron cloud only very slightly compressed by the surrounding carbon atoms. The cage serves to support the nitrogen structurally and protect it chemically so that it behaves like an almost perfectly isolated atom. This leads to the remarkable coherence of its spin.[29] The only other atom known to sit in a fullerene cage without bonding is phosphorous. P@$C_{60}$ shows electron spin properties similar to the incarcerated nitrogen, but for most purposes they are no better.[30,31] Since, of the two precursor gases used in the synthesis of these molecules, nitrogen is much easier to handle than phosphene, it is preferable to make N@$C_{60}$.[32]

The electron wavefunctions of N@$C_{60}$ are perturbed only weakly by the fullerene cage, so their energy levels in a magnetic field can to a good approximation be considered as those of an isolated nitrogen atom.[29] The $^{3/2}S$ electron states are split into four Zeeman levels, which in turn are each split into three hyperfine levels of $^{14}$N. To first order this gives three lines in the EPR spectrum. Higher order effects lift the degeneracy of the three $M_I = \pm 1$ transitions, leading to further splittings on the microtesla scale. Additional splittings due to hyperfine interactions with $^{13}$C nuclei distributed randomly on the $C_{60}$ cages give a measure of the interaction of the nitrogen electrons with the carbon cage. That the spectra can be resolved with sub-microtesla resolution is due to the exceptional degree of isolation of the nitrogen electron spins from the external environment, and the extraordinarily homogeneous environment of N@$C_{60}$ molecules in solution.

To maximise the spin flip lifetime $T_1$ and the superposition phase coherence time $T_2$ of N@$C_{60}$ it is necessary to minimise the adverse affects of the environment. To benefit fully from the natural almost complete absence of unwanted carbon nuclear spins, the fullerene molecules need to be in an environment which is also free from nuclear spins.[29] A good solvent for this purpose is $CS_2$, which has only a small concentration (0.76%) of sulphur isotope with non-zero nuclear spin. In investigating the temperature dependence of the spin properties, it is necessary to avoid grain boundary segregation of the fullerene molecules when the solvent freezes. For this purpose, a small amount of $S_2Cl_2$ can be added; the penalty in the additional nuclear spins being offset by the vitrifying effect.

8 January 2011

The spin relaxation time $T_1$ of N@C$_{60}$ increases progressively as the temperature is reduced.[29] Down to about 150 K it can be accounted for in terms of a two-phonon process which can be thought of as an Orbach mechanism. At lower temperatures $T_1$ does not increase as much as that would predict, possibly because of extra modes such as motion of the nitrogen atom within the fullerene cage. Nevertheless relaxation times approaching (and in dilute powder exceeding) a second have been observed, and there is every indication that longer times could be achieved at even lower temperatures. The spin coherence time $T_2$ has a more complicated behaviour. The outer coherences ($M_S = \pm 3/2 : \pm 1/2$) for $M \neq 0$ can be separately measured using the electron spin echo envelope modulation technique (ESEEM),[33] and for these $T_2$ falls below 1 µs below about 140 K, never to recover. This can be attributed to a zero field splitting whose fluctuations can no longer be averaged as the solvent becomes more viscous and eventually freezes. The inner coherences are not subject to zero field splitting, and their coherence times progressively improve at still lower temperatures, most likely limited by spectral diffusion with nuclear spins in the solvent. In principle the coherence times could therefore probably be increased by removing the small concentration of nuclear spins in the environment. However there is a better way to improve the memory time.

## 7. The nuclear spin as a resource

For electron spin states the presence of nuclear spins is often a nuisance. The absence of stable isotopes of any Group III or any Group IV elements which do not have nuclear spin has been a limitation for electron spin experiments in GaAs. Enormous ingenuity has gone into overcoming the decoherence caused by nuclear spins in these materials, with remarkable success. Magnetic nuclei have also been shown to dominate the relaxation mechanisms in molecular magnets at low temperatures.[34] But nuclear spins also provide a resource for quantum memory. The state of an electron qubit in a lithographically defined quantum dot has been stored in the surrounding nuclear spin ensemble.[35] It is even more convenient if there is a single nuclear spin coupled to the electron qubit.[36] Information which has been stored in the electron spin can be transferred to the nuclear spin, where it can be kept for much longer than the electron spin coherence time. A reversal of the process enables the information to be restored to the electron spin. For P donors in isotopically purified $^{28}$Si, storage times over 1 s have been demonstrated, and the fidelity of the recovered information confirmed by density matrix tomography. The principle has also been implemented in molecules of $^{15}$N@C$_{60}$, using the $I = ½$ spin of the $^{15}$N nucleus,[37] and should be applicable to any material system in which individual electron spins are coupled to an associated nuclear spin.

Just as nuclear spins are often regarded as a nuisance for long electron spin coherence times, so electron spins are generally regarded as undesirable for NMR experiments, including NMR for quantum information processing. But the same interactions which are used for nuclear spin memories can also be used for ultrafast control of nuclear spins. A technique known as *bang bang* control[38] can be applied to give immunity to external radiofrequency signals which would tend to change nuclear spin states; it also offers a mechanism for implementing phase gates on the nuclear qubit.[39] Starting with a nuclear spin superposition state, a microwave $2\pi$ pulse is applied at frequency that is selective of one component of the nuclear superposition. It might be thought that this would simply put the system back where it started, but, owing to the quantum nature of the spin of the nucleus, the nuclear superposition acquires a relative phase of π. Since it involves an electron spin transition, this operation is several orders of magnitude faster than an equivalent operation on the nuclear state that does not exploit the presence of the electron. The π phase shift inverts the evolution of the state of the nucleus in any external radiofrequency field, and rapid repetition of the microwave pulses (or *bangs*) decouples the nuclear spin from the external field entirely. Unlike the quantum Zeno effect, which maintains an eigenstate by a sequence of projective measurements, each of which has a finite probability of projecting the system into another state, *bang bang* control is fully deterministic, and can be used to maintain a state (within the limits of its $T_1$ and $T_2$) indefinitely in the presence of coherent or slowly varying external noise. Systematic errors in a single *bang* pulse are self-correcting if a *bang bang* sequence is used.

An electron spin in the vicinity can thus be a useful resource for systems in which nuclear spins are the primary information-carrying units. However, a disadvantage of the presence of the electron spin is that it can accelerate the decoherence of the nuclear spins. The best of both worlds can be achieved by using a photo-excited triplet state to polarize, couple and entangle to nuclear spins; once operations depending on the presence of the electron triplet are completed, the triplet is allowed to recombine, leaving no electron moments to cause dehoherence.[40] Proof-of-principle experiments using functionalised $^{13}$C labelled diethyl malonate (DEMF) have shown that a $^{13}$C nuclear spin can indeed be controlled in this way, and density functional theory modelling showed how the chromophore could be used to create controlled entanglement in a bis-DEMF molecule.

Controlled entanglement has been generated between electron and nuclear spins in an ensemble of P dopants in isotopically engineered $^{28}$Si.[41] At the temperature of 2.9 K and field of 3.4 T used, the nuclear spins would have been insufficiently polarized to establish true entanglement, and so the nuclear spins were hyperpolarized through a SWAP operation with the electron spins, followed by allowing the electron spins to rethermalize over a time short compared with the much longer spin relaxation time of the nuclear spins. A fidelity of 98% entanglement was achieved compared with the ideal state at that field and temperature. Combined with the single spin readout described in §5, conditions are now excellent for development of quantum computing technologies based on either single device[16] or cluster schemes.[42]

## 8. Charged NV⁻ centres in diamond

The nitrogen-vacancy (NV⁻) defect centre in diamond offers spin-selective optical transitions. It has a paramagnetic $S = 1$ ground state with a zero field splitting (lifting the degeneracy of the $M_S = 0$ and the $M_S = \pm 1$ levels), and it exhibits long coherence times (up to 1.8 ms in isotopically purified diamond at room temperature).[43] The centre is fluorescent, and the fluorescence intensity depends on $M_S$, thus allowing the optical detection of the spin state of single centres.[44] An intermediate metastable singlet state offers a route to dynamically polarising the spin in the $M_S = 0$ state. It will be extremely difficult to position NV⁻ centres at chosen locations with the atomic precision required to give reproducable interactions for quantum information processing, but coherent coupling been measured between two optically active centres which were about 10 nm apart and could be selectively addressed.[45] Spin-photon entanglement has been demonstrated from an NV⁻ centre.[46] Single spin coupling to light enables single spin measurement using the Faraday effect and unitary single spin manipulation using the optical Stark effect.[47] It may even be possible to use optical measurement to create entanglement between randomly placed centres which are sufficiently far apart that they do not interact.[48]

Since the host lattice for the NV⁻ centre is diamond, the natural abundance of carbon isotopes (1.1% $^{13}$C) ensures that a proportion of centres have a $I = ½$ $^{13}$C nucleus nearby, coupled to the electron by the hyperfine interaction. Thus using a combination of optical, microwave and radio-frequency excitations, it is possible to polarise and manipulate the $^{13}$C spin and then detect its state.[49] Sometimes multiple $^{13}$C nuclei can be found close to the NV⁻ centre and they can be selectively manipulated by virtue of their different couplings, thereby forming a small multi-qubit register.[50] A centre of this kind was used to generate examples of maximally entangled states of two $^{13}$C nuclear spins, with a further qubit on the electron spin to give three qubit entangled states.[51] This result stimulated lively discussion of how the quantum-information-carrying qubit states should be defined for a system of coupled spins,[52,53] an issue that has a bearing on the nature and degree of entanglement generated. The quality of debate illustrated just how subtle a phenomenon quantum entanglement is. It is quite different from anything familiar from the world of classical correlations, and if it can be harnessed it will open up utterly new applications.

## 9. Information storage in collective spin states

The coupling between a single electron spin and a single microwave photon is extremely weak. Using established principles of cavity quantum electrodynamics, the coupling can be enhanced by placing the electron in a resonant cavity. Nevertheless, assuming that the dimensions of the cavity are constrained by the wavelength in one dimension and by nanolithography in the other two dimensions, the coupling between a single microwave photon and a single spin is still very weak. The coupling between a single photon and the collective state of an ensemble of $N$ spins is stronger by a factor of $\sqrt{N}$. Superconducting microwave cavities suitable for integration with superconducting qubits can exhibit quality factors in excess of 5000 albeit at temperatures of a few tens of millikelvin.[54] Such cavities can also be coupled to electron spin ensembles, as has been demonstrated with $Cr^{+++}$ centres in ruby, and with N centres in diamond.[55] The coupling strength, which can be thought of as the rate with which photons are absorbed by the spin ensemble, was estimated to be up to 38 MHz, about six orders of magnitude greater than would be expected for a single spin in the cavity. In EPR experiments on $NV^-$ centres in diamond, the anticrossing gap was larger than the linewidths of both the superconducting cavity and the paramagnetic resonance, thus establishing the regime of strong coupling.[56]

The accessibility of this strong coupling regime raises the possibility of exploiting collective modes of the spin ensemble to store quantum information.[57, 58] A qubit encoded on a microwave photon in the resonator can be absorbed by the spin ensemble. The collective state of the spins that couples to the resonator mode, and which therefore is excited by the absorption of this photon, is characterised by each individual spin acquiring a small excitation with a uniform phase across the ensemble. Under a free evolution, this uniform-phase excitation of the spin ensemble will return to a single-photon excitation of the resonator, since the two modes are strongly coupled. However, if, when the excitation resides with the spins, a magnetic field gradient pulse is applied, the collective state of the spins acquires a phase that depends linearly on position within the ensemble, decoupling it from the resonator mode. The excitation, representing a qubit, is then locked into a mode of the ensemble characterised by a wavevector $k$. Applying a gradient pulse of the opposite polarity (or applying a π-pulse followed by a gradient pulse of the original polarity) restores the qubit to the uniform-phase $k = 0$ collective state, from where it can be read out into the resonator mode. In this way, it is possible to store multiple qubits within the ensemble in holographic modes that are orthogonal by virtue of their distinct wavevectors, and to recover them in arbitrary order with appropriate sequences of gradient pulses.

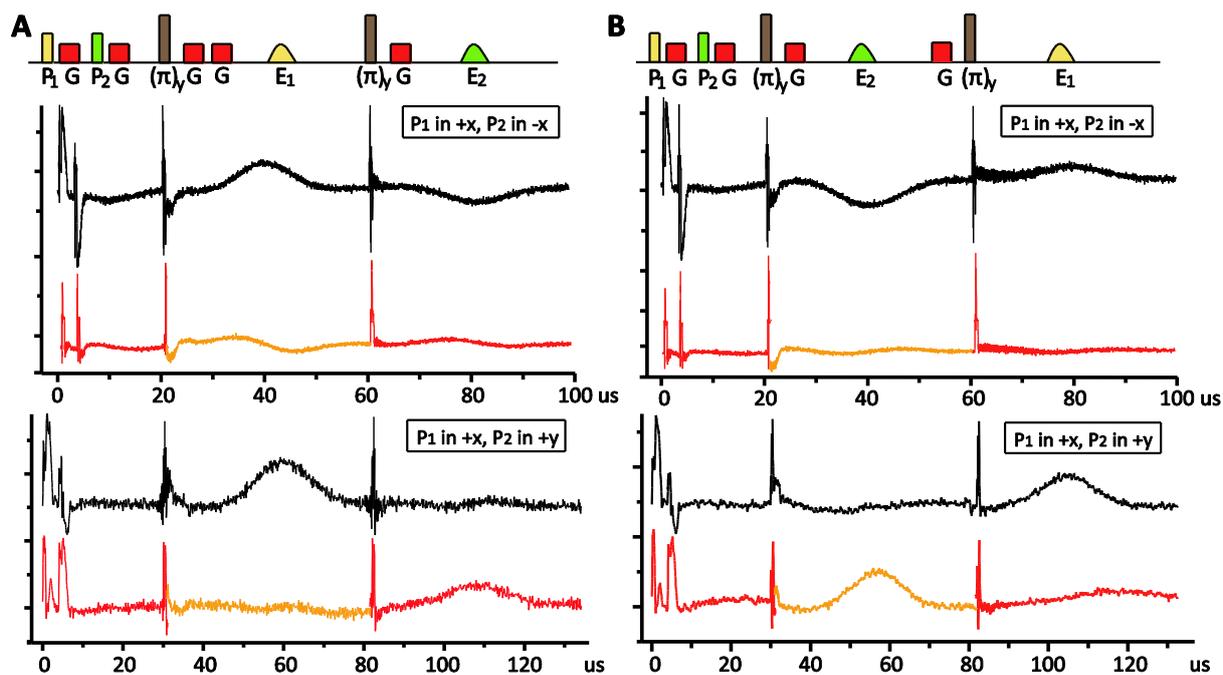

Figure 5: Recalling two stored microwave pulses in an arbitrary order using pulsed field gradient (A): Two microwave pulses recalled in the same order as they are stored. (B): Two pulses recalled in the inverse order. Transients are shown respectively for P1 and P2 applied in +x and −x, +x and +y direction. Each of the four diagrams shows the real part (+x direction, black) and imaginary part (+y direction, red) of the signal. As the refocusing pulse is in the y direction, the imaginary part of the signal between the first and second refocusing pulse (orange) is inverted to keep the echo and free induction decay (FID) having the same sign. In this experiment $P_1$ and $P_2$ are $\pi/6$ pulses. They are kept relatively strong so that we can observe a clear echo. Sample: $^{14}N@C_{60}$ at room temperature. From Ref. 59.

To demonstrate the principle, experiments have been performed using small microwave excitations involving many photons in a classical state, on spins in $N@C_{60}$ and phosphorous donors in silicon.[59] In $N@C_{60}$ at room temperature, the spin echo vanished when the excitation was transferred using a magnetic field gradient pulse into $k \neq 0$ mode. Figure 5 illustrates how two classical excitations of this kind stored in orthogonal modes could be read out in chosen order (thereby providing the basis for a random access memory). A further experiment using P:Si at 9 K demonstrated how up to 100 excitations may be stored and subsequently retrieved in reverse order. This represents the extension to electron spin systems of a demonstration, over 50 years ago, of the storage of 1000 excitations in a nuclear spin ensemble.[60] The collective excitation of the electron spin ensemble can be transferred to nuclear spins and subsequently restored to the electron spins and read out, using the same techniques as described in §7, thus offering a similar memory time.

Current superconducting qubits typically have coherence times measured in hundreds of nanoseconds, or at best microsceconds, and so far four Josephson junction devices have been connected together.[61] Entanglement between three transmon devices has been achieved in two different laboratories.[62, 63] A hybrid system, combining a small number of superconducting qubits (for multi-qubit logic gates) with a strongly-coupled electron spin ensemble (providing a large coherent qubit memory register), offers the appealing, albeit technically challenging, possibility of a quantum processor capable of operating on 100 or more qubits on the few-second time scale.[64]

## 10. Entanglement enhanced sensing

When a spin is in a superposition state in a magnetic field, the relative phase of the up and down states precesses at the Larmor frequency. Any change in field causes a change in the phase which accumulates in a given time, and this can provide the basis of magnetic field sensing. If an average is taken over $N$ spins, then the sensitivity to changes in the magnetic field can be enhanced by $\sqrt{N}$, giving rise to the standard quantum limit. However, if the $N$ spins can be prepared in the entangled state represented by the superposition of the states with all spins down and all spins up, then the energy difference between the two components of the superposition is $N$ times larger, and so the phase of the superposition is $N$ times more sensitive to changes in the magnetic field. The benefit of a further $\sqrt{N}$ in sensitivity over the standard quantum limit represents another powerful property of entangled states, and is known as the Heisenberg limit.

This has been demonstrated in an NMR experiment on trimethyl phosphite, as illustrated in Fig. 6.[65] The central phosphorous nucleus is surrounded by nine equivalent protons. A pulse which operates on the phosphorous spin puts it into a superposition state, and a second pulse then entangles all the protons with this nucleus and therefore with each other. The entangled state then accumulates a phase in an external magnetic field for a time $T_{wait}$, after which the process is reversed, transferring the phase acquired onto the phosphorous nucleus, which is then measured. The experiment is performed on a large ensemble of molecules, whose initial states are determined by thermal occupations of levels, at a temperature much higher than the Zeeman energy of the nuclear moments, so members of the ensemble exist with all possible arrangements for the spins of the protons. The spectral lines furthest from the zero of relative frequency in Figure 6 correspond to molecules that have the greatest number of proton spins aligned the same way; they give the smallest signals (because, statistically, there are fewest of them) but also the greatest amplification of phase when the field changes. The red curve shows the initial measurement with no phase change (effectively, $T_{wait} = 0$). The blue curve is taken with a 3.13 $\mu$T detuning of the magnetic field and $T_{wait} = 400\,\mu s$. As expected, the outermost spectral lines accumulate phase fastest, precessing at a rate 9.4 times faster than a single unentangled proton (the factor is more than 9, because the phosphorus nuclear spin participates in the entangled state). The experiment thus provides the basis for an order of magnitude improvement in sensitivity to magnetic fields over conventional proton-NMR-based sensors.

This experiment did not involve true entanglement, because for nuclear spins at practical temperatures and fields it is not possible to start with a pure state. Nevertheless the experiment demonstrates the phase amplification which would occur with fully entangled spins. Subsequent experiments using tetramethyl silane (in which a $^{29}$Si nuclear spin replaces the phosphorus as the central spin) have increased the number of spins to 12 + 1, have enhanced the populations of the states corresponding to the most sensitive outer lines, and eliminated errors in setting the $^{29}$Si frequency by decoupling that nucleus during $T_{wait}$.[66] Eventual applications of entanglement enhanced sensing may see the spins incorporated in a device for electrical detection of magnetic resonance. In order to evaluate such applications, it will be essential to develop the economic theory of how resources are to be counted relative to other techniques, and how the fundamental advantages of entanglement enhanced sensing depend on temperature, field, and size.[67]

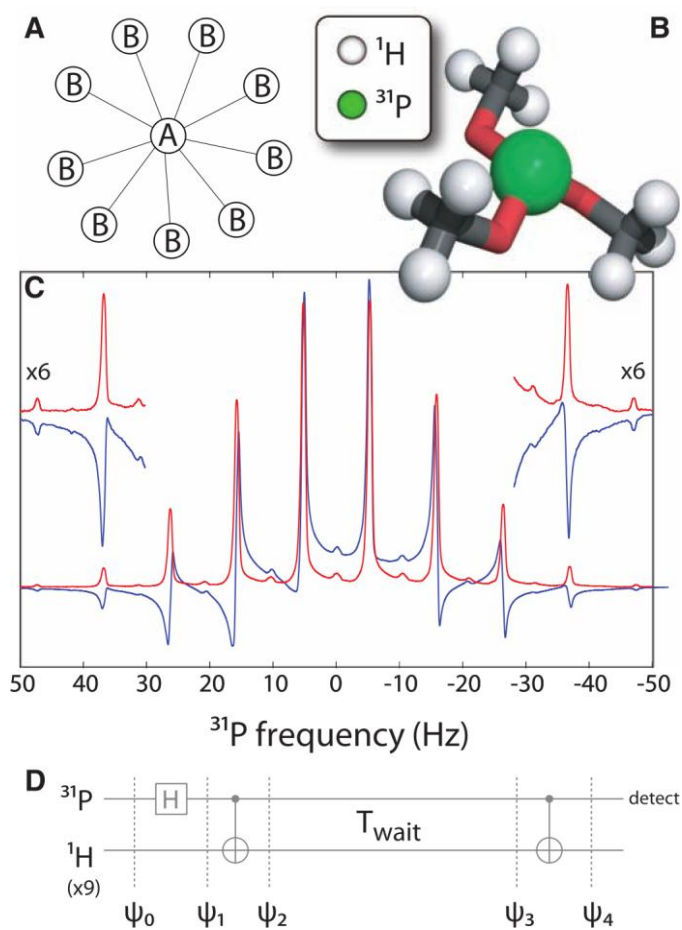

Fig. 6. Ten nuclear spins in a trimethyl phosphite (TMP) molecule. (**A**) Topology used to generate the spin state. (**B**) The TMP molecule consists of a central $^{31}$P nuclear spin surrounded by nine identical $^1$H spins. (**C**) The initial $^{31}$P NMR spectrum of TMP (red). Nuclear spin states were generated and allowed to evolve for 400 $\mu$s under detuning 3.13 $\mu$T. After mapping these entangled states back to the $^{31}$P, the resulting spectrum (blue) shows how the phase shift acquired increases for lines further from the centre. Low-intensity peaks between pairs of NMR lines arise from coupling to impurities. (**D**) Spin states were generated by first applying a Hadamard gate to the $^{31}$P followed by a C-NOT on the nine equivalent $^1$H. From Ref. 65.

## 11. Prospects for quantum spintronics

It is still too early to say which of the different materials systems considered here will be part of the winning technology for building a solid state quantum computer, but given their versatility and scope for interfacing with other degrees of freedom it is likely that electron spins will be at the heart of any solid state quantum technology.[68] Quantum dots in GaAs have provided the most mature fabrication technology for demonstrating the building blocks for spin-based manipulation and entanglement, at least at the two qubit level. Phosphorous in silicon has the enormous attraction of eventual compatibility with the vast silicon foundry industry, which will not easily be displaced in the foreseeable future for classical computers using CMOS, and the demonstration of nuclear memories enhances the information storage time by several orders of magnitude. Fullerenes have remarkable spin properties, and have provided a proving ground for some of the most advanced experiments on quantum control of single qubits in the form of spins. Once purified, all molecules are identical, and they offer the potential for chemical assembly of arrays and structural characterization by electron microscopy. NV$^-$ centres in isotopically purified diamond are also proving useable at room temperature, and they can be manipulated and read out optically. Each of these materials systems has its own strengths, and we can expect further rapid progress towards integration in electrically controlled devices. But a quantum computer may not be the first application of quantum spintronics. Entanglement enhanced sensors and

8 January 2011

metrology may find earlier applications, especially if the ensemble experiments described here can be translated into practical spintronic devices.

8 January 2011

8 January 2011